\documentclass[a4paper]{article}

\usepackage{ISCSLP2024}
\usepackage{cite}
\usepackage{graphicx}
\usepackage{makecell}
\usepackage{booktabs}
\usepackage{url}

\title{CUSIDE-array: A Streaming Multi-Channel End-to-End Speech Recognition System with Realistic Evaluations}
\name{
Xiangzhu Kong$^{1}$, Tianqi Ning$^1$, Hao Huang$^1$, Zhijian Ou$^{2}$
\thanks{
Corresponding author and project lead: Z. Ou. This work is supported by Guangxi Science and Technology Project (2022AC16002).
}
}
\address{
  $^1$School of Computer Science and Technology, Xinjiang University, Urumqi, China\\
  $^2$Speech Processing and Machine Intelligence (SPMI) Lab, Tsinghua University, China 
}
\email{
kongxiangzhu99@gmail.com, ozj@tsinghua.edu.cn
}

\begin{document}

\maketitle
\begin{abstract}
Recently multi-channel end-to-end (ME2E) ASR systems have emerged.
While streaming single-channel end-to-end ASR has been extensively studied, streaming ME2E ASR is limited in exploration.
Additionally, recent studies call attention to the gap between in-distribution (ID) and out-of-distribution (OOD) tests and doing realistic evaluations.
This paper focuses on two research problems: realizing streaming ME2E ASR and improving OOD generalization.
We propose the CUSIDE-array method, which integrates the recent CUSIDE methodology (Chunking, Simulating Future Context and Decoding) into the neural beamformer approach of ME2E ASR. It enables streaming processing of both front-end and back-end with a total latency of 402ms.
The CUSIDE-array ME2E models are shown to achieve superior streaming results in both ID and OOD tests. Realistic evaluations confirm the advantage of CUSIDE-array in its capability to consume single-channel data to improve OOD generalization via back-end pre-training and ME2E fine-tuning.  
\end{abstract}

\vspace{-0.1cm}
\section{Introduction}
\vspace{-0.1cm}

Multi-channel automatic speech recognition (ASR) systems have been continuously studied and improved, 
due to their ability to improve speech recognition robustness and accuracy through multi-channel inputs especially in far-field acoustic environments \cite{chime6, yu2022m2met, an2022exploiting}.
A beamforming front-end is usually introduced, before the ASR back-end, to leverage spatial information from multi-channel signals captured by an microphone array for speech enhancement \cite{arrays2001signal}.
Recently, neural beamformers have been developed, which integrate deep neural networks (DNNs) into classic signal processing based beamformers \cite{heymann2016neural}.

Conventionally, the beamforming front-end and the ASR back-end are optimized separately under different criteria \cite{heymann2016neural, xiang2016thu}. 
More recently, the multi-channel end-to-end (ME2E) ASR systems have emerged, which are interested in joint optimization of beamforming front-end and ASR back-end, by using the final ASR loss to optimize the entire system.
Some methods keep using neural beamformers and perform end-to-end training \cite{Beamnet_heymann, ochiai2017multichannel}.
Others, often referred to as all-neural, directly build multi-channel neural networks, which replace the classic beamformer with a multi-channel neural encoder to transform the multi-channel inputs \cite{concatenate, chang21_interspeech}.
It is expensive for the all-neural approach to exploit single-channel data, since simulated multi-channel data are needed, while it is cheap and effective for the neural beamformer approach by either back-end pre-training or data scheduling \cite{an2022exploiting}.
In order to further promote the application of ME2E ASR in the real world, this paper focuses on two research problems: realizing streaming ME2E ASR and measuring OOD generalization via realistic evaluations.

First, streaming ASR (a.k.a., online ASR) is of central importance in many real-world scenarios, whose goal is to emit recognition results as quickly and accurately as possible on the fly when the user is speaking.
While streaming methods for single-channel end-to-end ASR have been extensively studied (see Section 2 for more introduction), most studies in multi-channel end-to-end ASR are conducted and tested in full-context ASR (a.k.a., offline ASR). Most challenges in multi-channel ASR, such as ChiME \cite{chime6}, ASpIRE \cite{harper2015automatic} and the recent M2MeT \cite{yu2022m2met}, do not examine the streaming recognition capability.
Streaming processing of the front-end and back-end could be realized based on sliding windows (or say, blocks or chunks) or causal neural networks. 
For example, CGMM based online mask estimation has been studied in \cite{higuchi2016robust} for block beamformer.
Uni-directional LSTM network is used in \cite{matsui2018online} for DNN based online mask estimation.
Both still use full-context ASR back-end.
Streaming all-neural ME2E ASR could be built with causal attention layers \cite{chang21_interspeech}.
A recent block-based ME2E multi-talker ASR system in \cite{kanda2023vararray} introduces a total latency of 800ms.

In chunk-based streaming methods, using right contextual frames significantly improves recognition accuracy but bring additional undesirable latency. Recently, the CUSIDE framework (\underline{{C}}h\underline{{u}}nking, \underline{{Si}}mulating Future Context and \underline{{De}}coding) is proposed \cite{an22_interspeech}, which uses simulated future context and obtains state-of-the-art streaming ASR results.
In this paper, we propose the CUSIDE-array method, which integrates the CUSIDE methodology into the neural beamformer approach of ME2E ASR to enable streaming processing of both front-end and back-end with a total latency of 402ms.

To advance ME2E ASR in real-world applications, another important problem is how to evaluate and compare different systems. For system development, there usually exist some performance gap between in-distribution (ID) and out-of-distribution (OOD) tests, as pointed out in some recent studies \cite{taori2020measuring, radford2023robust}.
This paper aligns with this perspective of conducting both ID and OOD testings for realistic evaluations. In addition to evaluate and compare ID results, OOD testings are conducted, covering data from different benchmarks. These realistic evaluation results confirm the advantage of the CUSIDE-array method in its capability to consume single-channel data to improve robustness via back-end pre-training and ME2E fine-tuning\footnote{We release the code, scripts and data at the following URL \url{https://github.com/thu-spmi/CAT/blob/master/docs/cuside-array.md}.}.

\vspace{-0.1cm}
\section{Related work}
\vspace{-0.1cm}

\textbf{Multi-channel end-to-end ASR.~}
There are two main classes of methods for ME2E ASR: the neural beamformer based approach and the all-neural approach. 
% In neural beamformer based approach, the neural beamformer is cast as a differentiable component to allow end-to-end training of the beamformer and the ASR back-end under the ASR loss.
The neural beamformer could be designed by DNN based time-frequency mask estimation \cite{Beamnet_heymann,ochiai2017multichannel} or filter coefficient estimation~\cite{deep_beamforming_xiao,adaptive_beamforming}, within minimum variance distortionless response (MVDR) or generalized eigenvalue (GEV) beamformers.
The all-neural approach does not use an explicit beamformer, but designs a single neural network to learn the mapping of multi-channel inputs to ASR labels \cite{concatenate,chang21_interspeech,yu2023mfcca}. Recent progress includes introducing various new neural architectures, e.g., the cross-channel attention \cite{chang21_interspeech}, the multi-frame cross-channel attention and the multilayer convolutional mechanism to fuse the multi-channel output \cite{yu2023mfcca}.

\textbf{Streaming ASR.~}
Streaming methods have been extensively studied for single-channel end-to-end ASR models.
% the three classes of single-channel end-to-end ASR models, namely CTC~\cite{graves2006connectionist}, RNN-T~\cite{graves2012sequence} and attention-based encoder-decoder (AED) \cite{LAS}.
The general ideas are based on sliding windows (or say, chunks) or causal neural networks.
Chunk-based methods are attractive and employed in many previous studies, where non-causal networks (bi-directional LSTM or fully-connected self-attention) can be used for the chunk encoder \cite{SAA,CAT,MERL-DCN}, realizing full-context utilization in a chunk.
An important issue in chunk-based methods is that using right contextual frames can significantly benefit recognition accuracy but at the cost of latency. This issue is alleviated in the CUSIDE methodology, which uses simulated future context.
% For AED, additional efforts are needed to convert full sequence soft attention into local attention~\cite{mocha} in decoder.

\textbf{Realistic evaluations.~}
Recent studies call attention to the gap between ID and OOD tests, and the importance of realistic evaluations, which mean conducting both ID and OOD testings \cite{taori2020measuring, radford2023robust}.
ID testing (over held-out ID test sets) only measures ID generalization. 
It is important to compare how robust different models are to distribution shifts arising from real-world applications.
Robustness (or say, OOD generalization) is needed to be measured for a system.
In this paper, we conduct both ID and OOD testings for system comparisons, similar to evaluating the Whisper system \cite{radford2023robust}.
Training on larger and more diverse data is found to increase robustness \cite{taori2020measuring}.
This motivates us to opt for the neural beamformer approach that can effectively exploit richer single-channel data.

\begin{figure}[t!]
  \centering
  \setlength{\abovecaptionskip}{0pt}
  \includegraphics[width=\linewidth]{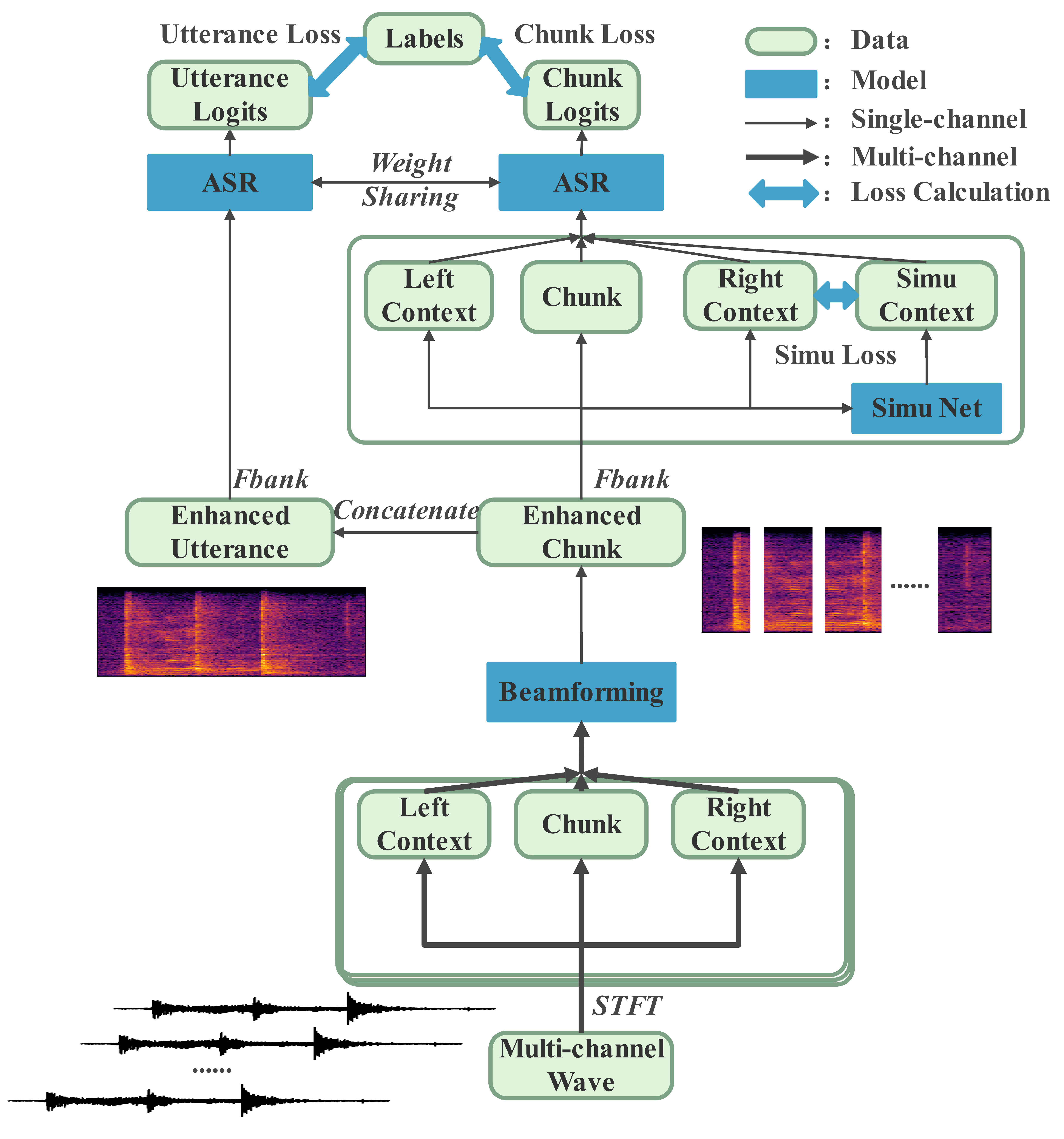}
  \vspace{-0.1cm}
  \caption{Overview of the CUSIDE-array method for streaming multi-channel end-to-end ASR. 
  For front-end chunking, right context of 0ms and real 400ms are randomized in training, while for back-end chunking, right context of 0ms, real 400ms, simulated 400ms are randomized in training. In both cases, no real future context is used in streaming recognition. For saving cost, we only use future context simulation in back-end chunking.
  }
  \label{fig:ME2E_ASR}
  \vspace{-0.5cm}
\end{figure}

\begin{table*}[h!]
    \caption{Results on AISHELL-4 test set.
    Exp 2 and 4 denote joint training of chunk-based streaming and non-streaming unified model.
    For Exp 1 and 3, no joint training means that whole-utterance models are trained with no chunking.
    The non-streaming recognition results are in parentheses.
    The right context (abbreviated as ctx) is not used by default during decoding. 
    Exp 5 and 6 denote that real and simulated right contexts are used in decoding respectively.
    Channel 0 is used for single-channel experiments. $\square$: not applied.
    }
    \vspace{-0.15cm}
    \label{tab:streaming_recognition}
    \setlength{\tabcolsep}{3pt}
    \centering
    \begin{tabular}{clccccc}
    \toprule
    \textbf{Exp} & \textbf{Model} & \textbf{Params (M)} & \textbf{\makecell{right ctx in\\training (ms)}} & \textbf{\makecell{right ctx in\\streaming recog. (ms)}} & \textbf{Latency (ms)} & \textbf{CER}\\
    \midrule
    1 & Single-channel E2E & 20.70 & $\square$ & 0 & 400 & 55.07 (38.76)\\
    2 & \hspace{2mm}+ joint training & 20.70 & 400 or 0 & 0 & 400 & 40.95 (36.17) \\
    \midrule
    3 & Multi-channel E2E & 25.77 & $\square$ & 0 & 400 & 56.84 (27.93) \\
    4 & \hspace{2mm}+ joint training & 25.77 & 400 or 0 & 0 & 400 & 36.68 (31.21) \\
    5 & \hspace{3mm}+ real right ctx (400ms) & 25.77 & 400 or 0 & 400 & 800 & 32.51 (31.21)\\
    6 & \hspace{3mm}+ simu right ctx (400ms) & 27.64 & 400 or 0 or [400]& [400] & 400 + 2 & 35.96 (31.70)\\
    \bottomrule
    \end{tabular}
    \vspace{-0.3cm}
\end{table*}

\vspace{-0.15cm}
\section{Method}
\vspace{-0.1cm}

The CUSIDE-array method is inspired by the CUSIDE methodology for streaming single-channel ASR \cite{an22_interspeech}, which introduces a simulation of future context for chunk-based streaming ASR.
In recognition, the CUSIDE-array method consists of context-sensitive chunking of multi-channel signals, chunk-based mask estimation and array beamforming, simulating future context from the single-channel enhanced speech, and ASR decoding, as overviewed in Figure~\ref{fig:ME2E_ASR}.

\vspace{-0.2cm}
\subsection{Context sensitive chunking}
\vspace{-0.1cm}
  
To enable streaming multi-channel ASR, we use context-sensitive chunking \cite{chen2016training,CAT} in both front-end and back-end.
First, multi-channel inputs are transformed into complex spectral features by Short-Time Fourier Transform (STFT) frame by frame. 
The utterance is then split into non-overlapping chunks.
For each chunk, a certain number of frames to the left and right of the chunk are spliced as contextual frames.
These spliced frames are collectively referred to as a context-sensitive chunk, which is fed into the front-end beamformer. The enhanced single-channel frames are then fed into the back-end ASR encoder.
Note that the output from the ASR encoder for these contextual frames is discarded in calculating the ASR loss.

\vspace{-0.1cm}
\subsection{Array beamforming}
\vspace{-0.1cm}
The front-end is basically a mask-based MVDR neural beamformer \cite{Beamnet_heymann,ochiai2017multichannel}. MVDR estimates the single-channel enhanced complex spectrum by applying a linear filter to the multi-channel STFT features. The filter coefficients are determined by the spatial covariance matrices of speech and noise signals, which are estimated from the time-frequency masks and the multi-channel STFT features.
In this work, a BLSTM network is run over context-sensitive chunks for mask estimation.

\vspace{-0.1cm}
\subsection{Future context simulation}
\vspace{-0.1cm}

Inspired by CUSIDE \cite{an22_interspeech}, a simulation network is introduced to recursively simulate the future contextual frames.
Specifically, a GRU-based unidirectional RNN is used to encode each new arriving chunk.
The GRU hidden state at the right boundary of the current chunk is then used to simulate the right context frames (the log Fbank features).
The simulation network is trained using an L1 based self-supervised loss, which is contained in the overall multi-tasking training, as detailed below.

\vspace{-0.1cm}
\subsection{Training}
\vspace{-0.1cm}
Similar to single-channel CUSIDE, the CUSIDE-array method performs multi-task training, which involves a simulation loss and joint training of the streaming model and the whole-utterance model which share their network parameters, with the following total loss:
$L_{total} = L_{utt} + L_{chunk} + \alpha L_{simu}$, 
where $L_{utt}$, $ L_{chunk}$, $ L_{simu}$ denote the non-streaming whole-utterance loss, the streaming chunk loss, and the simulation loss, respectively. $\alpha$ is the simulation loss weight.
In addition, chunk jittering (i.e., the chunk sizes are randomly draw from a uniform distribution, centering at the default size) and right context randomization (as explained in the caption of Figure 1) are also used in training \cite{an22_interspeech}.

\vspace{-0.1cm}
\begin{table*}[t]
    \caption{In-distribution (ID) and out-of-distribution (OOD) streaming and non-streaming results (in parentheses).
  ID results are underlined.
  E2E-FT refers to joint fine-tuning (FT) of the front-end (FE) and back-end (BE), with ID data (i.e. AISHELL-4) and simulated data for fine-tuning.
  Alimeeting-FE denotes the front-end from the ME2E CUSIDE-array model trained on Alimeeting.
MFCCA \cite{yu2023mfcca} adopts the all-neural approach of ME2E ASR, while CUSIDE-array belongs to the neural beamformer approach.}
  \vspace{-0.2cm}
    \label{tab:ID&OOD result}
    \centering
    \setlength{\tabcolsep}{3pt}
    \begin{tabular}{clc|cccc|c}
    \toprule
    &&& \multicolumn{5}{c}{Eval data / CER\% }\\
    \cmidrule(lr){4-8}
    \textbf{Exp} & \textbf{Model} & \textbf{Params (M)} & \textbf{AISHELL-4 test} & \textbf{Ali-test} & \textbf{Ali-eval} & \textbf{XMOS test} & \textbf{Average}\\
    \midrule
    2 & Single-ch. E2E (CUSIDE) & 20.70 & \underline{40.95 (36.17)} & 46.26 (41.23) & 50.10 (45.00) & 87.33 (86.34) & 56.16 (52.19)\\
    7 & \hspace{2mm}+ Pre-trained BE plug in& 80.72 & \underline{35.70 (26.41)} & 28.83 (20.29) & 29.07 (20.55) & 41.09 (29.80) & 33.67 (24.26) \\
    \midrule
    4 & Multi-ch. E2E (CUSIDE-array) & 25.77 & \underline{36.68 (31.21)} & 41.61 (36.21) & 45.27 (40.34) & 73.86 (66.24) & 49.36 (43.50) \\
    8 & \hspace{2mm}+ Pre-trained BE plug in & 85.79 & \underline{33.77 (20.27)} & 33.76 (17.94) & 34.46 (18.42) & 33.37 (22.57) & 33.84 (19.80)\\
    9 & \hspace{3mm}+ E2E-FT with ID (40h) & 85.79 & \underline{17.47 (14.22)} & 18.79 (14.52) & 20.22 (15.72) & 27.62 (17.92) & 21.03 (15.60)\\
    10 & \hspace{4mm}+ E2E-FT with simu (13h) & 85.79 & \underline{17.49 (14.14)}& 18.04 (13.83) & 19.11 (14.95) & 25.84 (20.69) & 20.12 (15.90) \\
    11 & \hspace{4mm}+ E2E-FT with simu (73h) & 85.79 & \underline{18.06 (14.46)} & 18.17 (13.65) & 19.11 (14.36) & 30.10 (21.19) & 21.36 (15.92)\\
    12 & \hspace{4mm}+ E2E-FT with simu (152h) & 85.79 & \underline{20.67 (14.62)} & 21.54 (13.93) & 22.26 (14.61) & 33.27 (21.39) & 24.44 (16.14)\\
    \midrule
    13 & \makecell[l]{Alimeeting-FE \\\hspace{2mm} + Pre-trained BE plug in} & 85.79 & 35.97 (21.76) & \underline{33.32 (17.90)} & \underline{34.98 (19.31)} & 35.84 (24.75) & 35.03 (20.93) \\
    \midrule
    14 & MFCCA (w/o LM) & 47.06 & \underline{$\square$ (21.69)} & \underline{$\square$ (12.80)} & \underline{$\square$ (13.97)} & $\square$ (61.79) & $\square$ (27.56)\\
    \bottomrule
    \end{tabular}
    \vspace{-0.3cm}
\end{table*}

\section{Experiments and results}
\vspace{-0.1cm}

\subsection{Data}
\vspace{-0.1cm}
AISHELL-4 \cite{fu2021aishell4} is used as the in-distribution (ID) data, which consists of 8-channel Mandarin meeting speech. 
As separation of overlapping speech is beyond the scope of this paper, the non-overlapping segments in the original training set are partitioned into the training and validation sets (43.4h and 2.3h respectively), and those in the original evaluation set are taken as the test set (8.9h).
We employed pyroomacoustics \cite{scheibler2018pyroomacoustics} for multi-channel data simulation, using settings similar to the ConferencingSpeech competition \cite{rao2021conferencingspeech}. 
The speech sources are from AISHELL-1 \cite{bu2017aishell1} and 60,000 hours of our in-house single-channel data. The noise sources, a total of 134 hours, are from MUSAN \cite{snyder2015musan} and CHiME-3 \cite{barker2017third}.

Three other testing sets are used to study OOD behaviors.
% \begin{itemize}
% \vspace{-0.13cm}
\textbf{Ali-test}, \textbf{Ali-eval}: non-overlapping speech segments in the test and evaluation set (3.6h and 1.2h respectively) from the Alimeeting dataset, which is the dataset used for the M2MeT challenge \cite{yu2022m2met}.
\textbf{XMOS test}: Real-world data captured by a 16-microphone rectangular array based on the XMOS solution\footnote{https://www.xmos.com/}, comprising approximately 40 utterances in a noisy environment. In testing, 10 channels were used, since the other 6 channels failed to record speech.
    % \vspace{-0.2cm}
% \end{itemize}

\vspace{-0.13cm}
\subsection{Experimental setup}
\vspace{-0.1cm}

The front-end utilizes a three-layer BLSTM network for mask estimation, with 320 hidden units in each layer's direction and a dropout rate of 0.5. Channel 0 is selected as the reference for MVDR beamforming and also for single-channel experiments. The CTC-based ASR encoder consists of a 12-layer Conformer model, which is configured with 4 attention heads, 256 attention dimension, and 3038 dimension in the feedforward layer.
The future context simulation network, following CUSIDE's design, is a three-layer unidirectional GRU with 256 hidden units, and 1 feed forward layer. The simulation loss weight $\alpha=0.975$.

For signal processing, we use 256-dimensional STFT features. 
For streaming processing, the chunk size is set to 400ms, with the left context of 800ms and the right context of 400ms.
The right context is randomly used in training, but never used in testing.
During training, the chunk size is randomly uniformly sampled from 350ms to 450ms, i.e., chunk jittering.
After beamforming, the enhanced STFT features are transformed into 80-dimensional log Fbank features.

Training employs the Adam optimizer and follows the Transformer learning rate scheduler \cite{vaswani2017attention}. Gradient clipping is applied to prevent divergence. The learning rate will decay with a factor of 0.1 if the loss does not decrease on the validation set, and training stops once it drops below $10^{-6}$. The final model averages the best 5 checkpoints according to the validation losses, while fine-tuning averages the last 3 checkpoints. In decoding, the language model (LM) is not used by default.

\vspace{-0.1cm}
\subsection{ID results with back-end trained from scratch}
\vspace{-0.1cm}

Different models are trained and evaluated on AISHELL-4's test set, measured by character error rate (CERs). The results are shown in Table \ref{tab:streaming_recognition} with the following observations.
1) It can be clearly seen that multi-channel models significantly outperform single-channel models (Exp 1 vs Exp 3, Exp 2 vs Exp4), demonstrating the significance of the front-end beamforming.
2) For Exp 1 and 3, whole-utterance models are trained with no chunking, which obtain good results in non-streaming recognition but with poor streaming results.
Joint training of chunk-based streaming and non-streaming unified model significantly improves the streaming results, in both single-channel and multi-channel experiments (Exp 1 vs Exp 2, Exp 3 vs Exp 4).
3) Examination of the results of Exp 4, 5, 6 reveals that using right context in decoding clearly improves streaming accuracy but introduces a delay.
Using simulated context improves accuracy with minimal delay (around 2ms, which is the run time cost for future context simulation on a GTX 1080 GPU).
The improvement of Exp 6 over Exp 4 in streaming ASR is significant, with p-value~=~4e-12 by matched-pair significance test \cite{significance}.

\vspace{-0.13cm}
\subsection{ID results with pre-trained back-end}
\vspace{-0.1cm}

A realistic problem in developing ME2E ASR system is how to exploit single-channel data, which is more abundant than multi-channel data. In CUSIDE-array, this can be realized by back-end pre-training.
We can then perform joint end-to-end fine-tuning of the front-end and the pre-trained back-end over AISHELL-4 training data, and examine the results on AISHELL-4 test set, as shown in Table ~\ref{tab:ID&OOD result}.
This is still a form of ID testing, but shows the superiority the CUSIDE-array method in exploiting single-channel data to enhance ID generalization. 

The pre-trained back-end is a CTC-based 12-layer Conformer network, trained on 60,000 hours of our in-house single-channel data. It includes 4 attention heads, 512 attention dimension, and 6600 feedforward layer dimension.

\begin{figure}[t]
  \centering
  \includegraphics[width=\linewidth]{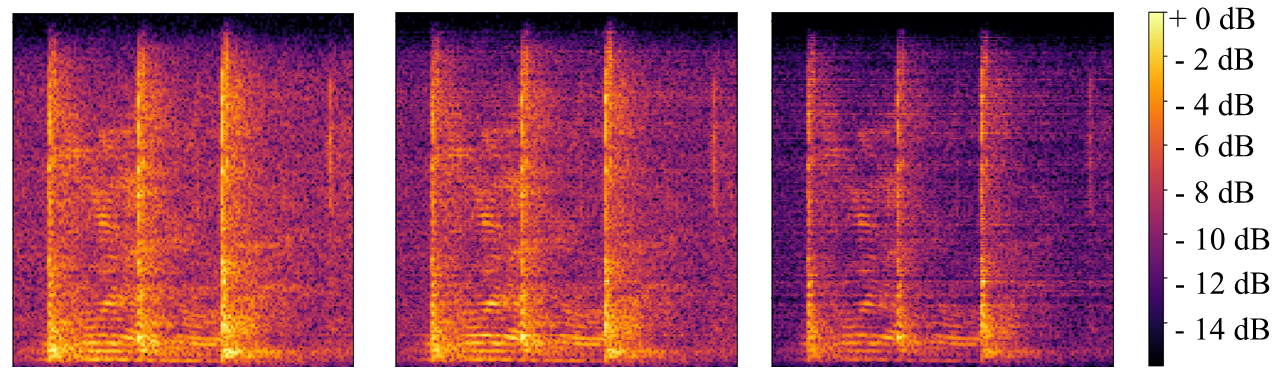}
  \caption{Comparison of original speech and enhanced speech spectrograms. Left: spectrogram of an original waveform of channel 0 in AISHELL-4. Middle/Right: spectrogram of the waveform after front-end enhancement in Exp 4 and Exp 9 respectively. ME2E fine-tuning (Exp 9) with a strong pre-trained back-end improves the front-end for better enhancement, obtaining less noisy spectrogram.}
  \label{fig:abc}
  \vspace{-0.5cm}
\end{figure}

First, we can see that plugging in the pre-trained back-end boosts the ID performance without any fine-tuning (Exp 2 vs 7, Exp 4 vs 8).
Then, comparing Exp 8 and 9, we can find that ME2E fine-tuning of the front-end and back-end with ID data (AISHELL-4) significantly further improve the ID performance.
Subsequently, we experimented with adding different amounts of simulated multi-channel data for further fine-tuning (Exp 10, 11 and 12) and observed that CER initially decreased and then slightly increased. 
This may reflect limited effect of adding multi-channel data simulated from single-channel speech for fine-tuning.
Finally, Figure \ref{fig:abc} shows the advantage of ME2E fine-tuning with a strong pre-trained back-end in improving the front-end performance.

\vspace{-0.13cm}
\subsection{OOD results with pre-trained back-end}
\vspace{-0.1cm}

This section shows the superiority the CUSIDE-array method in exploiting single-channel data via back-end pre-training and ME2E fine-tuning to enhance OOD generalization. 
Apart from AISHELL-4 test, three other testing sets are used to study OOD behaviors, as introduced in Section 4.1. For models trained/fine-tuned over AISHELL-4 training set, AISHELL-4 test is for ID testing, while others are for OOD testing.

For comparison, another two different models are tested.
One is the CUSIDE-array ME2E model trained on non-overlapping training data of Alimeeting; so for this model, Ali-test and Ali-eval represent ID testing.
The other is MFCCA \cite{yu2023mfcca}, a state-of-the-art model on Alimeeting.
MFCCA is a whole-utterance ME2E ASR model based on multi-frame cross-channel attention, adopting the all-neural approach. 
It is trained on a total of 917 hours of data from AliMeeting, AISHELL-4, and 600 hours of simulated overlapping speech. 
We use the code and model checkpoint released by the authors\footnote{https://github.com/alibaba-damo-academy/FunASR}. Its default configuration is used in our experiment.

Several research questions can be answered by checking Table ~\ref{tab:ID&OOD result}.
First, the simple concatenation of the ID front-end (from the CUSIDE-array model trained on Alimeeting) and the strong pre-trained back-end shows moderate results (Exp 13); its ID results (Ali-test, Ali-eval) are worse than the CUSIDE-array model fine-tuned on AISHELL-4 (Exp 9).
The comparison (Exp 13 vs Exp 9) reveals that ME2E fine-tuning of the front-end and pre-trained back-end with ID data (AISHELL-4) obtains superior OOD results on Ali-test and Ali-eval.
We can also see the same advantage of Exp 9 over Exp 13 for XMOS test (relative streaming CER reduction is 23\%), which are OOD testings for both models.
Second, compared with the CUSIDE-array model (Exp 9), MFCCA performs marginally better in its ID results (Ali-test, Ali-eval), but is far worse on OOD testings (XMOS test).
CUSIDE-array shows greater robustness to distribution shifts, while MFCCA exhibits a tendency to overfit to its ID results.
Third, compared to Exp 9 which is only fine-tuned on AISHELL-4, the OOD results of adding more simulated multi-channel data in ME2E fine-tuning are mixed, getting better in Ali-test, Ali-eval, but being worse in XMOS test. It seems that there is little transfer of robustness from synthetic to natural distribution shift, which is similar to the finding in \cite{taori2020measuring}. Training with simulated data does not necessarily improve robustness to real data.

\section{Conclusion and future work}
\vspace{-0.1cm}
In this paper, we propose the CUSIDE-array method, which integrates the recent CUSIDE methodology (using simulated future context) into streaming ME2E ASR.
The CUSIDE-array ME2E models are shown to achieve superior streaming results in both ID and OOD tests over AISHELL-4, Ali-test, Ali-eval, and XMOS test. 
Realistic evaluations confirm the advantage of CUSIDE-array in its ability to consume single-channel data to improve OOD generalization via BE pre-training and ME2E fine-tuning.
This paper is mainly concerned with multi-channel denoising. The integration of streaming dereverberation and separation within CUSIDE-array is interesting future work.

\bibliographystyle{IEEEtran}

\bibliography{mybib}

% \begin{thebibliography}{9}
% \bibitem[1]{Davis80-COP}
%   S.\ B.\ Davis and P.\ Mermelstein,
%   ``Comparison of parametric representation for monosyllabic word recognition in continuously spoken sentences,''
%   \textit{IEEE Transactions on Acoustics, Speech and Signal Processing}, vol.~28, no.~4, pp.~357--366, 1980.
% \bibitem[2]{Rabiner89-ATO}
%   L.\ R.\ Rabiner,
%   ``A tutorial on hidden Markov models and selected applications in speech recognition,''
%   \textit{Proceedings of the IEEE}, vol.~77, no.~2, pp.~257-286, 1989.
% \bibitem[3]{Hastie09-TEO}
%   T.\ Hastie, R.\ Tibshirani, and J.\ Friedman,
%   \textit{The Elements of Statistical Learning -- Data Mining, Inference, and Prediction}.
%   New York: Springer, 2009.
% \bibitem[4]{YourName17-XXX}
%   F.\ Lastname1, F.\ Lastname2, and F.\ Lastname3,
%   ``Title of your INTERSPEECH 2022 publication,''
%   in \textit{Interspeech 2022 -- 23\textsuperscript{rd} Annual Conference of the International Speech Communication Association, September 18-22, Incheon, Korea, Proceedings, Proceedings}, 2022, pp.~100--104.
% \end{thebibliography}

\end{document}